# Josephson penetration depth in coplanar junctions based on 2D materials

T. Li,[1,2,3,4,*] J. C. Gallop,[3] L. Hao,[3] and E. J. Romans[1,2,†]

[1]*London Centre for Nanotechnology, University College London, London WC1H 0AH, UK*

[2]*Department of Electronic and Electrical Engineering, University College London, London WC1E 6BT, UK*

[3]*National Physical Laboratory, Teddington TW11 0LW, UK*

[4]*QCD Labs and QTF Centre of Excellence, Department of Applied Physics, Aalto University, Espoo 02150, Finland*

[*]tianyi.li.14@ucl.ac.uk

[†]e.romans@ucl.ac.uk

## Abstract

Josephson junctions and SQUIDs with graphene or other 2D materials as the weak link between superconductors have become a hot topic of research in recent years, with respect to both fundamental physics and potential applications. We have previously reported ultra-wide Josephson junctions (up to 80 μm wide) based on CVD graphene where the critical current was found to be uniformly distributed in the direction perpendicular to the current. In this paper, we demonstrate that the unusually large Josephson penetration depth $\lambda_J$ that this corresponds to is enabled by the unique geometric structure of Josephson junctions based on 2D materials. We derive a new expression for the Josephson penetration depth of such junctions and verify our assumptions by numerical simulations.

# I. Introduction

Coplanar Josephson junctions and dc SQUIDs based on 2D materials with their critical current tunable by a local gate voltage have potential applications in low-noise superconducting electronics [1-17], especially in superconducting quantum circuits [18–20]. With graphene or other 2D materials as the weak link between two superconducting electrodes, Josephson junctions and dc SQUIDs can have their *I-V* characteristics easily tuned by the gate voltage, thanks to the low density of states in 2D materials. Recently we demonstrated Josephson junctions fabricated using Chemical Vapour Deposition (CVD) graphene and niobium superconducting electrodes [21]. The use of CVD rather than mechanically exfoliated graphene enables much more freedom for future device fabrication such as wafer-scale circuit design, but it also allowed us to study junctions as wide as 80 μm having easily measureable critical currents of order 10 μA at 1 K. We found that the critical current $I_c$ scaled linearly with the junction width *W*, and in an applied magnetic field *B*, an ideal Fraunhofer-like $I_c$-versus-*B* pattern was visible even for the widest junctions. This indicated that despite the unusually large junction width, the supercurrent was relatively uniformly distributed, and consequently that the Josephson penetration depth $\lambda_J$ must be comparable or even larger than this width.

The Josephson penetration depth is a key parameter for Josephson junctions, playing a similar role to the London penetration depth $\lambda_L$ for bulk superconductors. It arises because a large Josephson supercurrent can generate significant self-magnetic field, which affects the distribution of the gauge invariant phase difference $\varphi$ across the junction and in turn affects the supercurrent distribution via the first Josephson relation $j_s = j_c \sin \varphi$, where $j_c$ is the critical current density. For conventional low-$T_c$ tunnel junctions (Fig. 1(a)), the standard text book analysis [22–24] leads to a non-linear differential equation for $\varphi$ as a function of position *x* across the width of the junction of the form $d^2\varphi/dx^2 = (1/\lambda_J^2) \sin \varphi$ with

$$\lambda_J = \sqrt{\frac{\Phi_0}{2\pi\mu_0 j_c (L + 2\lambda_L)}}, \tag{1}$$

where $\Phi_0 \equiv h/2e$ is the magnetic flux quantum, $\mu_0$ is the permeability of free space, and *L* is the length of the junction (the thickness of the tunnel barrier). As a result, the phase decays from the edges of the junction towards the center over a length scale $\lambda_J$ which is the Josephson penetration depth. The ratio $\lambda_J/W$ determines the uniformity of the supercurrent and hence whether the critical current scales linearly with the width. Understanding this scaling is an important consideration in practical Josephson-based circuit design since device behavior is typically determined by the junction critical current amongst other parameters, and often it needs to be achieved fairly precisely.

The derivation of Eq. (1) assumes the flux in the junction corresponds only to the geometric inductance and ignores any kinetic inductance contribution. This assumption is not valid when the London penetration depth is comparable or larger than the superconducting electrode thickness, as might typically be the case for coplanar high-$T_c$ grain boundary junctions (Fig. 1(b)). In such junctions the kinetic inductance contribution typically dominates, and the Josephson penetration depth has instead been defined as [25]

$$\lambda_J = \sqrt{\frac{\Phi_0 W}{4\pi\mu_0 j_c \lambda_L^2}}, \qquad (2)$$

where $W$ is the width of the junction/electrode track.

Neither of the above expressions can be applied though to coplanar Josephson junctions based on 2D materials with low-$T_c$ electrodes such as niobium or aluminium (Fig. 1(c)). The electrodes are usually thick enough compared to $\lambda_L$ (Nb: 37.5 nm; Al: 19 nm) that the flux is dominated by the geometric inductance rather than the kinetic inductance, and so Eq. (2) is not applicable. Although Eq. (1) does correspond to the regime where geometric inductance dominates, it is not applicable to the coplanar geometry when we consider the actual current paths in the electrodes. This consideration leads us to introduce a new model yielding a different expression for the Josephson penetration depth of junctions based on 2D materials. According to the new expression, the Josephson penetration depth $\lambda_J$ is proportional to $1/\sqrt{t}$, where $t$ is the thickness of the 2D material. For coplanar Josephson junctions based on 2D materials, the Josephson penetration depth predicted by the new expression is much larger than substituting the same junction parameters into Eq. (1) or (2). We will also show that the Josephson penetration depth calculated by the new expression agrees well with the result of numerical simulation.

## II. Theoretical analysis

We consider coplanar Josephson junctions based on 2D materials using one of two possible alternative geometries shown schematically in Figs. 1(d) and 1(e). These represent limiting cases: in Geometry A, the width of the superconducting electrodes is much larger than the width of the 2D material; while in Geometry B, the width of the superconducting electrodes is the same as that of the 2D material. Most of the experiments in the literature adopted Geometry A [1,3,6,8,9,12,14,17,18,26,27], while a few of them are better described by Geometry B (or somewhere between Geometry A and Geometry B [2,4,10,11,13,15,16,19,20]). When the superconducting electrodes are thicker than the London penetration depth $\lambda_L$, the supercurrent tends to flow in a narrow region within $\sim \lambda_L$ from the edge of the superconducting electrodes as shown. In the 2D material, the distribution of the supercurrent is more uniform, depending on the Josephson penetration depth $\lambda_J$, which is the theme of this paper.

For a sandwich-shaped tunnel junction (Fig. 1(a)) carrying a Josephson current in the $z$-direction, the derivation of Eq. (1) starts from the assumption that the electrodes are thick enough in the $y$-direction to exclude flux from their bulk, and relates the flux $B_y(L + 2\lambda_L)dx$ linking a closed rectangular integration loop in the barrier region extending $\lambda_L$ into each electrode between $x$ and $x + dx$, to the change in the gauge invariant phase difference $\varphi(x + dx) - \varphi(x)$ between opposite sides of the loop. Further differentiation leads to

$$\frac{d^2\varphi}{dx^2} = -\frac{2\pi(L + 2\lambda_L)}{\Phi_0}\frac{\partial B_y}{\partial x}, \tag{3}$$

from which Eq. (1) follows by writing $\partial B_y/\partial x$ in terms of the Josephson current using the 4$^{th}$ Maxwell equation, and assuming $\partial B_x/\partial y$ can be neglected for an such an infinitely thick junction. Equation (3) is still valid for 2D junctions with either Geometry A or B; however, we cannot find a simple expression for $\partial B_y/\partial x$ because the thickness of the junction in the $y$-direction is now negligible. So instead we consider the supercurrent to flow in an infinitely thin surface in the $xz$ plane, and calculate the magnetic field at a given point within the junction by integrating the contributions from all parts of the supercurrent. For convenience we divide the supercurrent flow into several regions as indicated on Fig. 1. Region 1 is the flow within the 2D material, along the length of the junction; Region 2 is the flow on the edge of the superconducting electrodes in direct contact to the 2D material, along the width of the junction; and Region 3 is the flow on other edges of the superconducting electrodes, parallel or perpendicular to the length of the junction depending the geometry considered. Regions 1 and 2 are the same for both geometries; the only difference between the two geometries is Region 3.

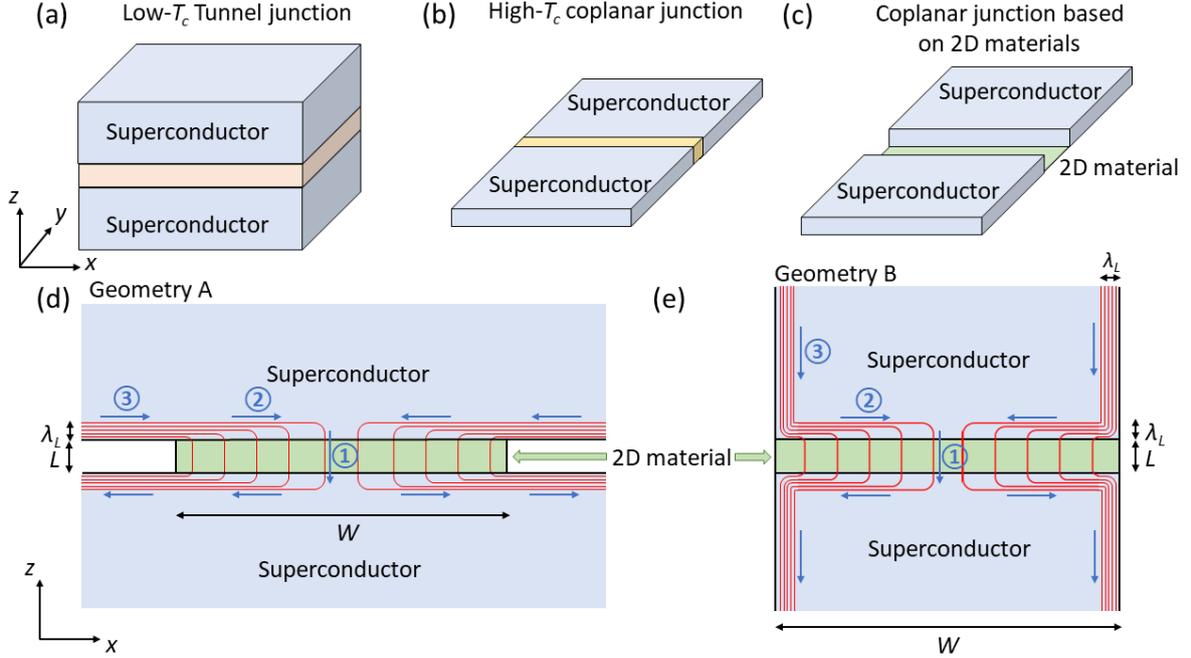

FIG. 1. Schematic diagrams of different types of Josephson junctions. (a) A conventional tunnel junction. (b) A high-$T_c$ coplanar junction. (c) A coplanar junction based on 2D materials. (d)(e) Top view of two geometries of Josephson junctions based on 2D materials. In both geometries, the 2D material lies parallel to the plane of the page, between two thin-film superconducting electrodes. In Geometry A, the superconducting electrodes can be regarded as infinitely wide; while in Geometry B, the superconducting electrodes are of the same width $W$ as the 2D material, and extend to infinity in the $z$-direction. The red lines display the distribution of supercurrent in the superconducting electrodes and the 2D material. In the superconducting electrodes, the supercurrent is distributed mainly within the width of the London penetration depth $\lambda_L$ from the edges of the superconductors. The blue arrows indicate the direction of the supercurrent. ①, ②, and ③ indicate the regions of supercurrent considered individually in the main text when calculating the magnetic field generated by the supercurrent.

In Part I, we mentioned the interrelationship between the distribution of supercurrent and the distribution of the magnetic field. On the one hand, the supercurrent generates a contribution to the total magnetic field; on the other hand, the magnetic field determines the distribution of the gauge invariant phase difference $\varphi$ which itself determines the distribution of supercurrent. Therefore, Eq. (3) has to be solved self-consistently by some numerical technique to obtain the actual distribution of supercurrent. However to derive an approximate analytical expression for $\lambda_J$, we make an assumption that for typical devices, the junction width is likely comparable or smaller than the Josephson penetration depth, so that the supercurrent is approximately uniform, with a supercurrent density of $j_{2D}$ (per unit width). We will revisit this assumption in Part III. Since the junction length $L$ is much smaller than the junction width $W$, we assume the magnetic field in the 2D material is independent of the coordinate $z$. We first consider the magnetic field $B_y(x)$ in the $y$-direction at a given point $x$ in the 2D material that is generated by all line elements of Region 1. The lengths of the line elements are taken as $L + 2\lambda_L$, which extend into the superconducting electrode by $\lambda_L$ on each side. According to the Biot-Savart law,

$$B_{1y}(x) = -\int_{-W/2}^{W/2} \frac{\mu_0 j_{2D}(L+2\lambda_L)}{4\pi(x-x')\sqrt{(x-x')^2 + (L+2\lambda_L)^2/4}} dx'$$

$$= \frac{\mu_0 j_{2D}}{2\pi}\left[\arctan\frac{\sqrt{(x-W/2)^2 + (L+2\lambda_L)^2/4}}{(L+2\lambda_L)/2} - \arctan\frac{\sqrt{(x+W/2)^2 + (L+2\lambda_L)^2/4}}{(L+2\lambda_L)/2}\right]. \quad (4)$$

Secondly, we consider the magnetic field in the $y$-direction at a given point $x$ in the 2D material that is generated by the supercurrent in Region 2. For simplicity, this is taken to be flowing at a distance of $\lambda_L$ from the superconductor edge (($L/2 + \lambda_L$) from the centre of the 2D material). Again, applying the Biot-Savart law,

$$B_{2y}(x) = -\int_{-W/2}^{W/2} \frac{\mu_0 j_{2D} x'(L+2\lambda_L)}{4\pi[(x-x')^2 + (L+2\lambda_L)^2/4]^{3/2}} dx'$$

$$= \frac{\mu_0 j_{2D}}{\pi(L+2\lambda_L)}\left[\frac{x^2 - Wx/2 + (L+2\lambda_L)^2/4}{\sqrt{(x-W/2)^2 + (L+2\lambda_L)^2/4}} - \frac{x^2 + Wx/2 + (L+2\lambda_L)^2/4}{\sqrt{(x+W/2)^2 + (L+2\lambda_L)^2/4}}\right]. \quad (5)$$

Next, we consider the magnetic field $B_{3ay}(x)$ in the y-direction at a given point $x$ in the 2D material that is generated by the supercurrent in Region 3 in Geometry A. This is taken to be flowing at a distance of $\lambda_L$ from the superconductor edge, to infinity in the ±x-direction, so

$$B_{3ay}(x) = -\frac{\mu_0 j_{2D} W}{2\pi(L + 2\lambda_L)} \left[ \frac{x - W/2}{\sqrt{(x - W/2)^2 + (L + 2\lambda_L)^2/4}} \right. \\ \left. + \frac{x + W/2}{\sqrt{(x + W/2)^2 + (L + 2\lambda_L)^2/4}} \right]. \tag{6}$$

Similarly, we consider the magnetic field $B_{3by}(x)$ in the y-direction at a given point $x$ in the 2D material that is generated by the supercurrent in Region 3 in Geometry B. This is taken to be flowing on the superconductor edges, to infinity in the ±z-direction, so

$$B_{3by}(x) = -\frac{\mu_0 j_{2D} W}{4\pi} \left[ \frac{2x}{x^2 - W^2/4} - \frac{(L + 2\lambda_L)/2}{(x + W/2)\sqrt{(x + W/2)^2 + (L + 2\lambda_L)^2/4}} \right. \\ \left. - \frac{(L + 2\lambda_L)/2}{(x - W/2)\sqrt{(x - W/2)^2 + (L + 2\lambda_L)^2/4}} \right]. \tag{7}$$

In Geometry A, the total magnetic field in the y-direction in the 2D material is then $B_{1y} + B_{2y} + B_{3ay}$; whereas in Geometry B it is $B_{1y} + B_{2y} + B_{3by}$. For a Josephson junction based on 2D materials with Nb as the superconducting electrodes and graphene as the weak link, we can choose $L = 50$ nm, $W = 80$ μm, $\lambda_L = 37.5$ nm, and a value of critical current $j_{2D} = 0.1$ A/m corresponding to our previously fabricated devices [21]. Using these parameters, we calculate the variation of $B_{1y}$, $B_{2y}$, $B_{3ay}$, and $B_{3by}$ across the junction, as shown in Fig. 2(a). For $B_{3by}$, the model is over-simplified at the boundaries $x = \pm W/2$ (since we have ignored the exact distribution of the supercurrent in the x- and y-directions as $x \to \pm W/2$), so consequently it predicts $B_{3by} \to \infty$. Except for the narrow regions within $\sim \lambda_L$ of these boundaries, the total magnetic field in either geometry is dominated by $B_{2y}$, which changes almost linearly with $x$. However since $\lambda_L \ll W$, we may assume the supercurrents in the narrow regions close to the boundaries make negligible contribution to the total supercurrent of the junction. Therefore, except for these narrow regions, we can write

$$\frac{\partial B_y}{\partial x} \approx \frac{\partial B_{2y}(x=0)}{\partial x} = -\frac{\mu_0 j_{2D}}{\pi(L+\lambda_L)}\frac{W}{\sqrt{W^2/4+(L+2\lambda_L)^2/4}} \approx -\frac{2\mu_0 j_{2D}}{\pi(L+2\lambda_L)}. \quad (8)$$

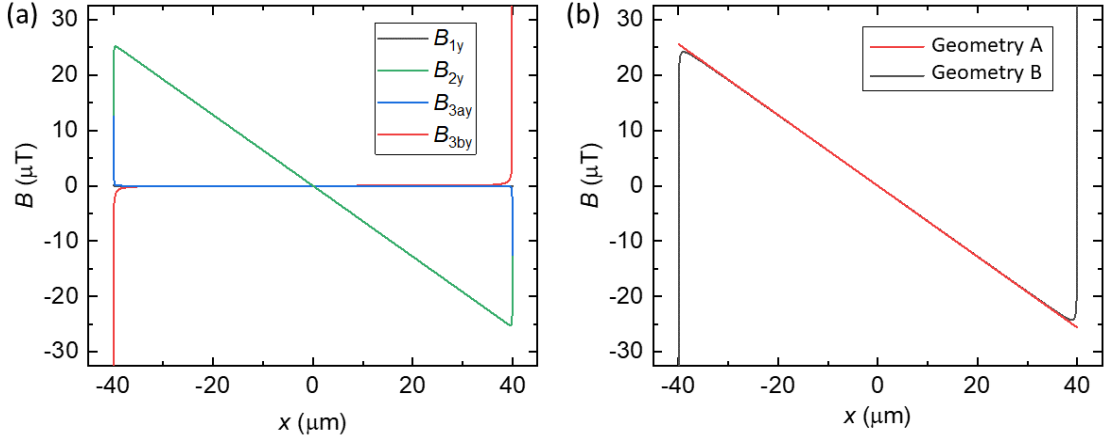

FIG. 2. Variation of the magnetic field in the *y*-direction in the 2D material generated by the supercurrent across the junction width for parameters given in the main text. (a) The magnetic field generated by different regions of the supercurrent as marked in Fig. 1. $B_{1y}$ is much smaller than others that it is hardly visible. (b) The total magnetic field in Geometry A and B respectively. In Geometry A, the total magnetic field is $B_{1y} + B_{2y} + B_{3ay}$; in Geometry B it is $B_{1y} + B_{2y} + B_{3by}$. In both geometries, the total magnetic field is dominated by $B_{2y}$.

Substituting this into Eq. (3) gives

$$\frac{d^2\varphi}{dx^2} = -\frac{2\pi(L+2\lambda_L)}{\Phi_0}\frac{\partial B_y}{\partial x} \approx \frac{4\mu_0 j_{c,2D} W}{\Phi_0}\sin\varphi, \quad (9)$$

where $j_{c,2D}$ is the critical current density per unit width. Similar to how the conventional Josephson penetration depth is defined in Eq. (1), we propose that for Josephson junctions based on 2D materials, the Josephson penetration depth should be

$$\lambda_J = \sqrt{\frac{\Phi_0}{4\mu_0 j_{c,2D}}} = \sqrt{\frac{\Phi_0}{4\mu_0 j_c t}}, \quad (10)$$

where $t$ is the thickness of the 2D material (single or multiple atomic layer thickness), and $j_c$ is the critical current density per unit cross-sectional area. Comparing Eq. (10) with Eq. (1), we can see that for conventional tunnel junctions, $\lambda_J \propto 1/\sqrt{L + 2\lambda_L}$; while for coplanar Josephson junctions based on 2D materials, $\lambda_J \propto 1/\sqrt{t}$. For coplanar Josephson junctions based on single- or few-layer 2D materials, $t \ll L + 2\lambda_L$, so the Josephson penetration depth given by Eq. (10) should be much larger than the value calculated by Eq. (1). Note that Eq. (10) results from the unique structure of 2D junctions: (i) the 2D material is so thin that the distribution of supercurrent in the *y*-direction is considered as a δ-function; and (ii) the superconducting electrodes are thicker than the London penetration depth so that the kinetic inductance can be ignored and Eq. (3) is still valid.

For a Josephson junction based on mono-layer graphene, $t = 0.3$ nm. So for a critical current density $j_{c,2D} = 0.1$ A/m, the Josephson penetration depth predicted by Eq. (10) is 63 μm. However, if we use the conventional expression, Eq. (1), the Josephson penetration depth is only 2.4 μm. Consequently Eq. (10) better matches the length scale implied by the experimental results in Ref. [21].

## III. Numerical simulations

To obtain an analytical expression of the Josephson penetration depth in Part II, we assumed the junction width is smaller than the Josephson penetration depth so that the supercurrent is still almost uniformly distributed across the width of the junction. Another assumption we made is that the singularities in $d^2\varphi/dx^2$ at the boundaries of the junction have negligible effect on the distribution of supercurrent. In this part, we validate these two assumptions by numerical simulations.

To avoid the first assumption in the numerical simulations, we use an iterative process to numerically solve

$$\frac{d\varphi}{dx} = -\frac{2\pi(L + 2\lambda_L)}{\Phi_0}(B_{ext} + B_{sc}), \tag{11}$$

where $B_{ext}$ is the external magnetic field and $B_{sc}$ is the magnetic field generated by the supercurrent. In such a way, we consider the interrelation between the distribution of supercurrent and the distribution of magnetic field self-consistently.

To avoid the second assumption in the numerical simulation, the step size along the width of the junction is chosen to be much smaller than the London penetration depth $\lambda_L$, so that we can make sure that the effect of the supercurrent on the edges of the junction is taken into consideration.

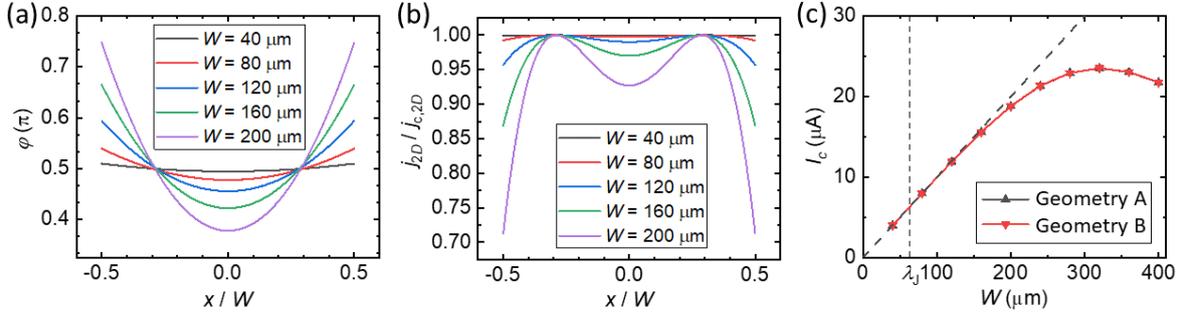

FIG. 3. The simulated distributions of (a) the gauge invariant phase difference $\varphi$ and (b) the supercurrent density $j_{2D}$ across the width of the junction for junctions with different widths, when there is no external magnetic field and when the critical current is reached. The other junction parameters are given in the main text. The curves shown here are for Geometry A, however, there is hardly any noticeable difference for Geometry B. (c) The critical current versus the junction width. As the junction width exceeds the Josephson penetration depth, the critical current does not increase linearly with the junction width.

We numerically simulated junctions with different widths from 40 μm to 200 μm using the same junction parameters as in Part II. We first consider the case without external magnetic field. The distributions of the gauge invariant phase difference $\varphi$ and of the supercurrent density $j_{2D}$ across the width of the junction are plotted in Fig. 3. As can be seen in Fig. 3(a), the critical current (maximum of the supercurrent) is reached when $\varphi$ is distributed as close to $\pi/2$ as possible. When the junction width is smaller than or comparable with the Josephson penetration depth (which is 63 μm according to Eq. (10)), the deviation of $\varphi$ from $\pi/2$ is relatively small; as the junction width becomes larger than the Josephson penetration depth, the deviation of $\varphi$ from $\pi/2$ becomes more and more significant. The supercurrent density is related to $\varphi$ by the first Josephson relation and plotted in Fig. 3(b). When the junction width is smaller than or comparable to the Josephson penetration depth, the supercurrent density is almost uniform and close to the critical current density; as the junction width becomes larger than the Josephson penetration depth, the supercurrent becomes less and less uniform and shows two peaks in the distribution. We have simulated the distributions of $\varphi$ and $j_{2D}$ for both Geometries A and B, and there is hardly any noticeable difference in the distributions. The distribution of supercurrent density in Fig. 3(b) indicates that as the junction width increases the critical current of the whole junction does not scale linearly with the junction width like $I_c = j_{2D}W$. As shown in Fig. 3(c), the critical current reaches its maximum when the junction width is about five times the Josephson penetration depth predicted by Eq. (10).

We further consider the distribution of the critical current under an external magnetic field. For Josephson junctions of finite widths but much narrower than the Josephson penetration depth, the critical current $I_c$ will show an ideal Fraunhofer-like pattern under an external magnetic field due to self-interference. In the simulation, we assume that the external magnetic field is evenly distributed across the effective area of the junction, $A_{eff} = (L + 2\lambda_L)W$. In Fig. 4, we plot the simulated critical current versus the magnetic flux inside the effective area, $\Phi = BA_{eff}$, for junctions with different widths in Geometry A and B respectively. For both geometries, when the junction width is smaller than, or comparable to, the Josephson penetration depth given by Eq. (10), the interference pattern is quite similar to an ideal Fraunhofer-like pattern. When the junction width much exceeds the Josephson penetration depth, the interference pattern becomes distorted from an ideal Fraunhofer-like pattern in the following three ways: (i) the maximum critical current at zero external magnetic field becomes significantly smaller than the critical current density times the junction width, as shown in Fig. 3(c); (ii) the first few minima of the critical current on both sides of the central peak do not reach zero, which means the supercurrent components within the junction no longer cancel with each other under those magnetic fields; and (iii) there appear some small peaks near the first few minima of the critical current. Although the numerical simulation here cannot tell us the exact value of the Josephson penetration depth, it does show that the Josephson penetration depth predicted by Eq. (10) is in the right order of magnitude, and confirm the validity of the two assumptions in the derivation of Josephson penetration depth for junctions based on 2D materials.

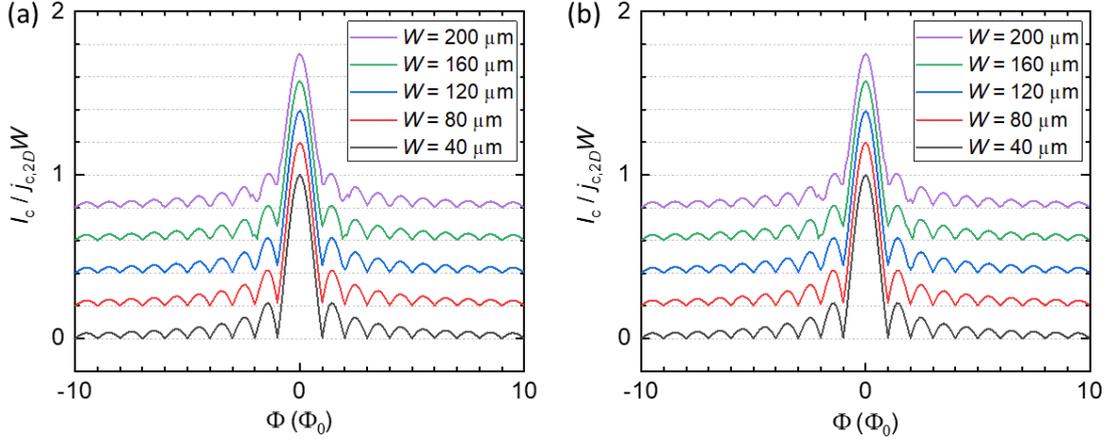

FIG. 4. The simulated self-interference pattern in an external perpendicular magnetic field for junctions with different widths. (a) Geometry A; (b) Geometry B. As the junction width becomes larger than the Josephson penetration depth $\lambda_J = 63$ μm, the ideal Fraunhofer-like pattern becomes more and more distorted. The curves are offset along the vertical axis by steps of 0.2 for clarity.

## IV. Conclusion

In conclusion, we have proposed a new expression for the Josephson penetration depth of coplanar junctions based on 2D materials. The Josephson penetration depth is a crucial parameter for designing such junctions, since it sets the limit of where the critical current scales with the junction width. We have shown that the Josephson penetration depth of this sort of junction is proportional to $1/\sqrt{t}$, and is around tens of micrometres for typical Josephson junctions based on graphene, much wider than most conventional tunnel junctions. That means even for junctions as wide as tens of micrometres, the supercurrent is relatively uniformly distributed across the width of the junctions. Such ultra-wide and uniform junctions could allow us to fabricate multiple local gates across the width of a single junction, with their positions and shapes engineered so that we can arbitrarily control the local critical current density, thus controlling the total critical current and even realising novel self-interference patterns within a single junction under an external magnetic field.

## Acknowledgment

The work was funded by the UK National Measurement System (NMS) awards 119616 and 119610, by the EU project EMPIR 17FUN06 SIQUST, and by the UK Engineering and